%% file: conference_101719.tex
\UseRawInputEncoding
\documentclass[conference]{IEEEtran}
\usepackage{placeins}
\IEEEoverridecommandlockouts
\usepackage{cite}
\usepackage{amsmath,amssymb,amsfonts}
\usepackage{algorithmic}
\usepackage{graphicx}
\usepackage{tabularx}
\usepackage{textcomp}
\usepackage{subcaption}
\usepackage{xcolor}
\usepackage{booktabs} 
\usepackage{algorithm} 
\usepackage{array}
\usepackage{multirow}
\usepackage{array}
\usepackage{tabularx,booktabs,siunitx,caption}
\newcolumntype{C}{>{\centering\arraybackslash}X}
\def\BibTeX{{\rm B\kern-.05em{\sc i\kern-.025em b}\kern-.08em
    T\kern-.1667em\lower.7ex\hbox{E}\kern-.125emX}}
\begin{document}

\title{An Efficient and Explainable KAN Framework for Wireless Radiation Field Prediction}

\author{\IEEEauthorblockN{Jingzhou Shen, Xuyu Wang}
\IEEEauthorblockA{
Knight Foundation School of Computing and Information Sciences, Florida International University, Miami, FL 33199, US\\
Emails: jshen020@fiu.edu, xuywang@fiu.edu}
}

\maketitle
\input{sec/0_abstract}    
\input{sec/1_intro}

\input{sec/2_Preliminary}

\input{sec/3_model}

\input{sec/4_experiments}

\input{sec/5_conclusion}

\bibliographystyle{IEEEtran}
\bibliography{ref}
\end{document}

%% file: sec/0_abstract.tex
\begin{abstract}
Modeling wireless channels accurately remains a challenge due to environmental variations and signal uncertainties. Recent neural networks can learn radio frequency~(RF) signal propagation patterns, but they process each voxel on the ray independently, without considering global context or environmental factors. Our paper presents a new approach that learns comprehensive representations of complete rays rather than individual points, capturing more detailed environmental features. We integrate a Kolmogorov-Arnold network (KAN) architecture with transformer modules to achieve better performance across realistic and synthetic scenes while maintaining computational efficiency. Our experimental results show that this approach outperforms existing methods in various scenarios. Ablation studies confirm that each component of our model contributes to its effectiveness. Additional experiments provide clear explanations for our model's performance.
\end{abstract}


%% file: sec/1_intro.tex
\section{Introduction}
\label{sec:intro}
Wireless channel modeling and prediction are essential for communication systems, including network planning, localization, and antenna design. Traditional approaches mainly fall into two categories: statistical models and ray-tracing simulations~\cite{almers2007survey}. Statistical models capture average signal behavior but lack geometric details, limiting their spatial and temporal resolution. Ray tracing simulates individual signal paths using 3D environment models, producing site-specific predictions by calculating each ray's parameters and summing their contributions. However, ray tracing requires intensive computation and detailed 3D maps, limiting its practicality in complex environments. Both approaches treat multipath components individually, lacking a unified representation that captures spatial variations and interactions among components.

Traditional machine learning methods have made significant  wireless channel modeling. For data generation, Sionna RT~\cite{10465179} provides a differentiable ray tracer for radio propagation modeling, while Jaensch et al.~\cite{jaensch2024radio} introduce simulated path loss radio maps using realistic city maps. In applying machine learning, PMNet~\cite{lee2024scalable} uses an encoder-decoder structure for accurate channel prediction, and RadioUNet~\cite{9354041} generates pathloss estimations using UNet architectures that transfer to real-world scenarios.

The emergence of 3D vision technologies like neural radiance fields~(NeRF)~\cite{mildenhall2021nerf} has created new opportunities for modeling 3D environments. These approaches benefit fields such as computer graphics, augmented reality, and scene reconstruction by capturing spatial relationships. Wireless channel modeling is well-suited for 3D vision techniques because radio signals travel through space and interact with the environment through reflection, diffraction, and scattering similar to the light transport phenomena modeled by NeRF.

Recent research has applied data-driven 3D vision techniques to wireless channel modeling. NeRF$^2$~\cite{nerf2} integrates electromagnetic wave propagation physics into the learning framework, but it requires substantial data. NewRF~\cite{lu2024newrf} addresses this by using direction-of-arrival (DOA) as prior knowledge to learn from sparse measurements. For generalization to unseen scenes, GWRF~\cite{yang2025gwrfgeneralizablewirelessradiance} employs a geometry-aware transformer combined with neural-driven ray tracing. Moreover, RFCanvas~\cite{10.1145/3666025.3699351} utilizes visual priors to adapt to environmental changes.

\begin{figure}[htbp]
    \centering
    \begin{subfigure}[b]{0.5\linewidth}
        \centering
        \includegraphics[width=\linewidth]{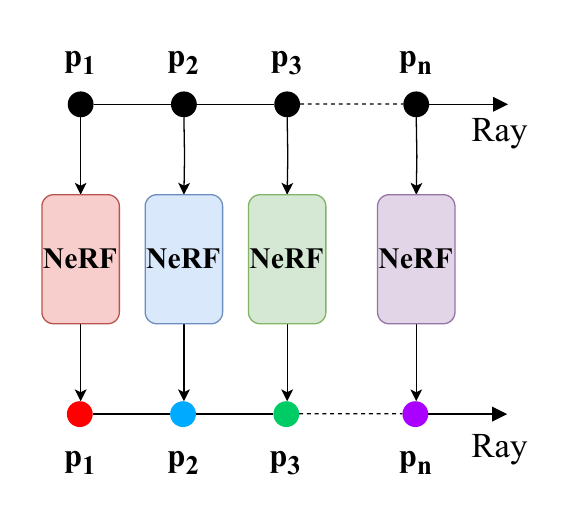}
        \caption{NeRF: point-in-point-out.}
        \label{p1}
    \end{subfigure}%
    \hfill%
    \begin{subfigure}[b]{0.5\linewidth}
        \centering
        \includegraphics[width=\linewidth]{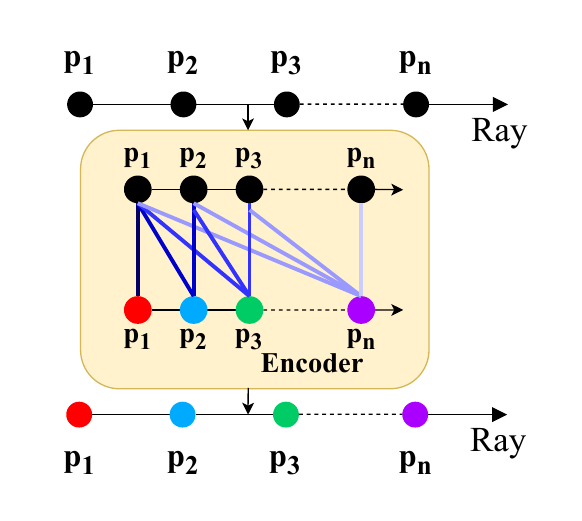}
        \caption{Ours: ray-in-ray-out.}
        \label{r1}
    \end{subfigure}
    \caption{Comparison between the traditional NeRF and our approach.}
    \label{fig:differ}
\end{figure}

Despite recent advances in NeRF-based wireless modeling, significant challenges still exist. Most current methods analyze points along rays independently, missing important interactions between ray segments that are essential for representing complex phenomena such as multipath propagation. While the traditional point-in-point-out model, illustrated in Fig.~\ref{p1}, processes each point independently to predict channel features, it fails to capture spatial dependencies along the ray path. In contrast, we propose a ray-in-ray-out model, shown in Fig.~\ref{r1}, which encodes the entire ray as a sequence of sampled points. This approach enables the network to learn interactions along the ray path and develop detailed spatial representations while aligning with physical wireless signal propagation behavior, resulting in more accurate modeling of spatially correlated channel features.

To address these challenges, we propose KANNA-NeRF: \textbf{KAN} with li\textbf{N}ear \textbf{A}ttention \textbf{NeRF}, which processes the entire ray as input to learn high-dimensional scene representations. Our approach integrates a Kolmogorov-Arnold network (KAN)~\cite{liu2025kan} with a transformer architecture~\cite{vaswani2017attention} to better encode data and contextual relationships, while providing result explainability. The main contributions of this paper are as follows.

\begin{itemize}
    \item We propose a ray-in-ray-out encoder that replaces the traditional point-to-point model in the NeRF framework to incorporate contextual information between ray points, enabling the model to capture more comprehensive relationships and information.
    \item We implement masked attention mechanisms to encode high-dimensional features in alignment with ray-tracing characteristics and use KAN instead of multi-layer perceptrons (MLPs) for enhanced representational capacity.
    \item Our method demonstrates superior performance when tested on both real-world and simulated wireless communication scenarios, surpassing key baseline approaches. We also provide explanatory visualizations demonstrating how our results correspond to real-world conditions.
\end{itemize}

%% file: sec/2_Preliminary.tex
\section{Preliminary}
\label{sec:Preliminary}
In this section, we provide background for wireless channel prediction and the standard NeRF applied to wireless propagation analysis, as well as the introduction of the KAN model.

\subsection{Wireless Channel Prediction}
\label{subsec:WCP}
A wireless communication system consists of a transmitter (TX) and a receiver (RX). The transmitter converts information into a modulated signal for transmission by manipulating amplitude and phase to encode data onto a carrier wave. The transmitted signal is represented as a complex number $X = A e^{j\theta}$, where $A$ denotes amplitude and $\theta$ represents phase. During propagation, the signal experiences amplitude reduction due to path loss and attenuation, while its phase changes due to multipath propagation and transmission distance variations. The attenuation factor $\Delta A$ accounts for amplitude scaling caused by energy loss during propagation. The phase rotation parameter $\Delta \theta$ represents the cumulative phase shift induced by the transmission environment. Consequently, in a free-space path loss model, the received signal at RX can be expressed as follows,
\begin{equation}
    Y = A e^{j\theta} \times \Delta A e^{j\Delta\theta} = A \cdot \Delta A e^{j(\theta + \Delta\theta)}.
\end{equation}

Electromagnetic wave propagation in wireless environments is affected by physical phenomena such as reflective surfaces, absorptive materials, diffractive edges, and penetrable objects. When signals meet an obstruction, these obstacles act as secondary emission points, distributing electromagnetic energy throughout the environment. Due to these wave interactions, the signal at the receiver is a superposition of multiple replicated waveforms from the original transmission. This multipath phenomenon can be expressed mathematically as:

\begin{equation}
Y = A e^{j\theta} \times \sum_{l=0}^{L-1} \Delta A^l e^{j\Delta\theta_l},
\end{equation}
where $L$ represents the total number of propagation paths, and $l$ is the path index. 

A wireless channel represents the physical medium through which signals travel from transmitter to receiver. It accounts for essential phenomena including path loss, fading, and multipath propagation. The channel can be expressed as follows,
\begin{equation}
H = \frac{Y}{X} = \sum_{l=0}^{L-1} \Delta A^l e^{j\Delta\theta_l}.
\end{equation}

\subsection{NeRF for Wireless Scenes}
\label{subsec:SNAWS}
Vanilla NeRF processes three key inputs: the positions of sampled points along the ray and the transmitter, as well as the direction of a ray from the receiver. This can be modeled as:
\begin{equation}
\begin{aligned}
F_{\Theta} : &\; (\underbrace{\mathbf{x, y, z}}_{\text{Voxel Position}}, 
                  \underbrace{\alpha, \beta}_{\text{Measuring Direction}}, \quad \underbrace{\mathbf{x}_{\text{tx}}, \mathbf{y}_{\text{tx}}, \mathbf{z}_{\text{tx}}}_{\text{Transmitter Position}}) \\
            &\longrightarrow 
              (\underbrace{\delta}_{\text{Attenuation}}, 
               \underbrace{\xi}_{\text{Signal Emission}}).
\end{aligned}
\end{equation}

\begin{figure*}[htbp]
\centering
\includegraphics[width=\textwidth]{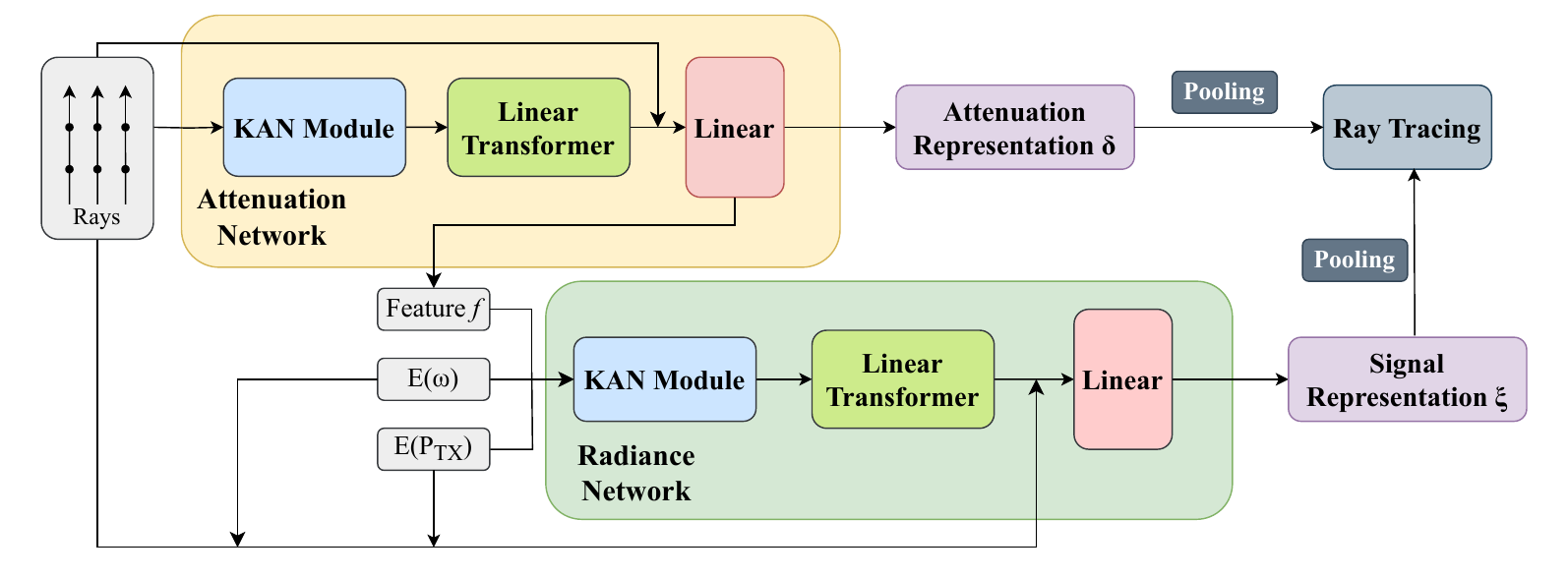}
\caption{KANNA-NeRF framework based on NeRF$^2$.}
\label{figure_nu}
\end{figure*}

Specifically, the output ($\delta$, $\xi$) represents the electromagnetic wave properties within each volumetric element. Here, $\delta$ indicates signal strength attenuation and $\xi$ represents signal emission. Each spatial element acts as an independent signal origin point, consistent with the Huygens-Fresnel principle of wave propagation~\cite{Born_Wolf_Bhatia_Clemmow_Gabor_Stokes_Taylor_Wayman_Wilcock_1999}.

Vanilla NeRF conducts electromagnetic ray tracing using neural network-based scene representations. For a receiver at position $P_\text{RX}$, the signal received from direction $\omega$ is calculated by integrating all signals emitted by voxels along that ray:
\begin{equation}
P(r, \omega) = P_{\text{RX}} + r \cdot \omega,
\end{equation}
where $r$ is a scalar that progresses along the ray from the RX to a maximum distance $D$. The voxels $\{ V_1, \dots, V_S \}$ are sampled along this ray. Each voxel functions as a secondary emission source and is processed by the NeRF to produce the outputs described above. The retransmitted signal at each voxel is derived by combining signals from preceding voxels, attenuated by intervening voxels, and then emitted through the radiance term $\xi$ of the current voxel. Thus, the received signal from direction $\omega$ can be expressed as:
\begin{equation}
R(\omega) = \sum_{i=1}^{S} \exp
\underbrace{
  \left(\sum_{j=0}^{i-1}\delta_j\right)
}_{\text{Attenuation}}
\; \cdot \; 
\underbrace{\xi_i}_{\text{Signal Emission}}.
\end{equation}

\subsection{KAN}
The KAN model is a neural network architecture recently developed to overcome certain limitations inherent in traditional neural networks, particularly MLPs. Traditional MLPs utilize predefined and fixed activation functions, such as rectified linear units (ReLU) or Gaussian error linear units (GELU). Although these fixed activations are effective for many tasks, they often fail to capture complex, nonlinear relationships. This limitation can lead to underfitting or require unnecessarily large networks with extensive parameter adjustments. Moreover, the rigid form of these fixed functions provides limited interpretability regarding how input variables contribute to the final predictions.

KAN is introduced to address these shortcomings by replacing fixed-form activations with learnable, data-adaptive activation functions. Specifically, each neuron in a KAN layer computes its output as a sum of nonlinear transformations of its input features, where these transformations are parameterized as spline functions. Formally, given an input vector $x \in \mathbb{R}^{d}$, the output of a KAN layer, $y \in \mathbb{R}^{m}$, is calculated as:
\begin{equation}
    y_i = \sum_{j=1}^{d} f_{ij}(x_j), \quad i = 1,2,\dots,m,
\end{equation}
where $f_{ij}$ represents a single-variable spline-based activation function individually learned for each input dimension $x_j$.

The spline activation function employed by KAN is typically implemented using a set of piecewise polynomial basis functions, such as cubic B-splines. Each spline function $f_{ij}$ is parameterized by a set of learned knots (control points) and corresponding spline coefficients:
\begin{equation}
    f_{ij}(x_j) = \sum_{k=1}^{K} a_{ijk} B_{k}(x_j),
\end{equation}
where $K$ is the number of spline basis functions, $\{a_{ijk}\}$ represents the spline coefficients during training using gradient-based methods, and $\{B_k(\cdot)\}$ denotes the basis functions defined by the knot positions. KAN can learn flexible, highly expressive nonlinear mappings directly from the data by adjusting knot locations and spline coefficients dynamically.

%% file: sec/3_model.tex
\section{Framework}
\label{sec:Model}

This section presents our KANNA-NeRF model structure shown in Fig.~\ref{figure_nu}. We use a ray-in-ray-out design~\cite{wang2022next} to better capture dependencies within the model. Unlike the point-in-point-out scheme in NeRF, which predicts each sample independently, our method considers the ray as a continuous entity. The ray-in-ray-out approach is well suited for wireless channel modeling, where signals naturally propagate along ray paths. It provides a more detailed and reliable characterization for high-fidelity wireless channel prediction.

As illustrated in Fig.~\ref{figure_nu}, our model consists of two main components: an attenuation network and a radiance network. The attenuation network outputs the attenuation representation that quantifies how much electromagnetic signal is reduced along each ray segment. These outputs describe the exponential decay of the signal due to absorption and scattering in the environment. Conversely, the radiance network outputs scattered radiance signal representations. These outputs allow the model to capture multipath reflections and specular highlights that cannot be explained solely by attenuation. Together, the two networks provide a unified volumetric rendering formulation for wireless channel prediction.


\subsection{CSI Data Pre-processing}
Prior to model training, we apply a pre-processing pipeline to channel state information~(CSI) datasets to address signal distortions in uplink channel estimation. When base stations estimate channels from client preamble symbols, distortions arise from carrier frequency offset~(CFO) and hardware detection delay. While these distortions cannot be fully eliminated, they can be standardized through a three-step transformation inspired by FIRE~\cite{10.1145/3447993.3483275}.

First, we implement signal strength scaling by normalizing the channel matrix $H$ using $H_{\text{norm}} = \frac{\sqrt{N}}{\|H\|}H,$ where $N$ denotes the number of base station antennas and $\|H\|$ is the Frobenius norm of $H$. This normalization ensures energy consistency across antennas, aligning with conservation of power principles. Second, we eliminate the CFO impact by dividing the channel matrix by the value measured at the first antenna, effectively removing global phase rotations while preserving relative channel relationships. This follows from the fact that CFO introduces a uniform phase shift across antennas, which can be factored out without altering spatial diversity. Third, we mitigate hardware detection delay by applying linear regression to estimate the phase slope across subcarriers for the first antenna, then subtracting this slope from all antennas. This step compensates for group delay distortions, which are theoretically modeled as a linear phase term in the frequency domain.

Although synthetic data is free from random real-world noise, it can still contain systematic artifacts introduced by simulation inaccuracies, rendering engines, or channel modeling assumptions. Our CSI pre-processing thus provides a standardized representation consistent with electromagnetic propagation theory: normalization enforces power stability, CFO correction ensures phase coherence, and delay compensation preserves frequency-domain linearity. These transformations reduce domain gap effects, improve the statistical alignment between synthetic and real-world wireless measurements, and enhance the generalization of models trained on both domains.

\subsection{The Attenuation Network}
The attenuation network comprises three essential components: a PowerMLP module, which is an efficient version of KAN, a linear Transformer component, and a linear layer with the residual connection. Each component plays a critical role in modeling how electromagnetic signals attenuate as they propagate through the environment.

\subsubsection{KAN Module}
First, we implement the PowerMLP as our efficient KAN module. Research by Coffman~\cite{coffman2025matrixkanparallelizedkolmogorovarnoldnetwork} indicates that KAN modules are computationally intensive, especially when using high-degree B-splines. Additionally, Qiu et al.~\cite{qiu2024powermlp} demonstrate that KAN networks typically require ten times more training time than standard MLPs with equivalent parameters due to the iterative nature of spline computations. Therefore, we implement the PowerMLP approach~\cite{qiu2024powermlp} to replace traditional MLPs, combining training efficiency with high expressiveness. The key innovation in PowerMLP is replacing spline-based activation functions with power-of-ReLU functions in the KAN framework, which offers non-iterative and faster computation. These power-of-ReLU functions maintain the expressiveness of splines while significantly reducing computational overhead. The PowerMLP substitution can be understood by examining how spline activations decompose into simpler functions. The KAN model uses computationally demanding spline-based activations defined by,
\begin{equation}
S(x) = \sum_{m} \alpha_{m} B^{(L)}_m(x),
\end{equation}
where $\alpha_m$ are the learnable coefficients and $B^{(L)}_m(x)$ represents the B-spline basis functions of order $L$. In fact, PowerMLP demonstrates that any $L$-order B-spline can be expressed as a linear combination of shifted ReLU$^{L-1}$ functions:
\begin{equation}
\rho_\ell(x) = \left(\max\{0,x\}\right)^\ell,
\end{equation}
with spline coefficients incorporated into the weights. This approach significantly reduces computational cost while avoiding iterative evaluations.

The attenuation network should learn a rich, spatially varying field of absorption and phase shifts across many voxels. Each layer refines how material properties and scene density combine to attenuate an incoming wave, so a deeper stack is needed to capture complex interactions and long-range correlations in the medium. We use 8 layers of PowerMLP as our KAN module here.  

From a memory complexity perspective, PowerMLP offers clear advantages over conventional KANs while remaining comparable to standard MLPs. For an $L$-layer network with hidden width $d$ and batch size $B$, the parameter storage scales as $\mathcal{O}(L \cdot d^2),$ identical to that of a standard MLP. The additional cost of storing activations during training is $\mathcal{O}(L \cdot B \cdot d),$ again on par with MLPs, with only a minor constant factor increase due to the power-of-ReLU evaluations. In contrast, KANs with spline degree $k$ require $\mathcal{O}(L \cdot d^2 \cdot k)$ for parameters and $\mathcal{O}(L \cdot B \cdot d \cdot k)$ for activations, often making them an order of magnitude more memory-intensive. Thus, PowerMLP achieves the expressive benefits of KANs while keeping the spatial complexity essentially the same as MLPs, enabling efficient large-scale training.

\subsubsection{Linear Transformer Module}
Then, we apply a linear transformer module to process the features from the PowerMLP. For computational efficiency, we implement Performer\cite{choromanski2021rethinking} as our attention mechanism to aggregate context from points along the ray. Traditional transformer networks face scalability challenges when processing long sequences, such as those encountered in dense ray sampling for wireless channel modeling.

\begin{figure}[htbp]
\centering
\includegraphics[width=\linewidth]{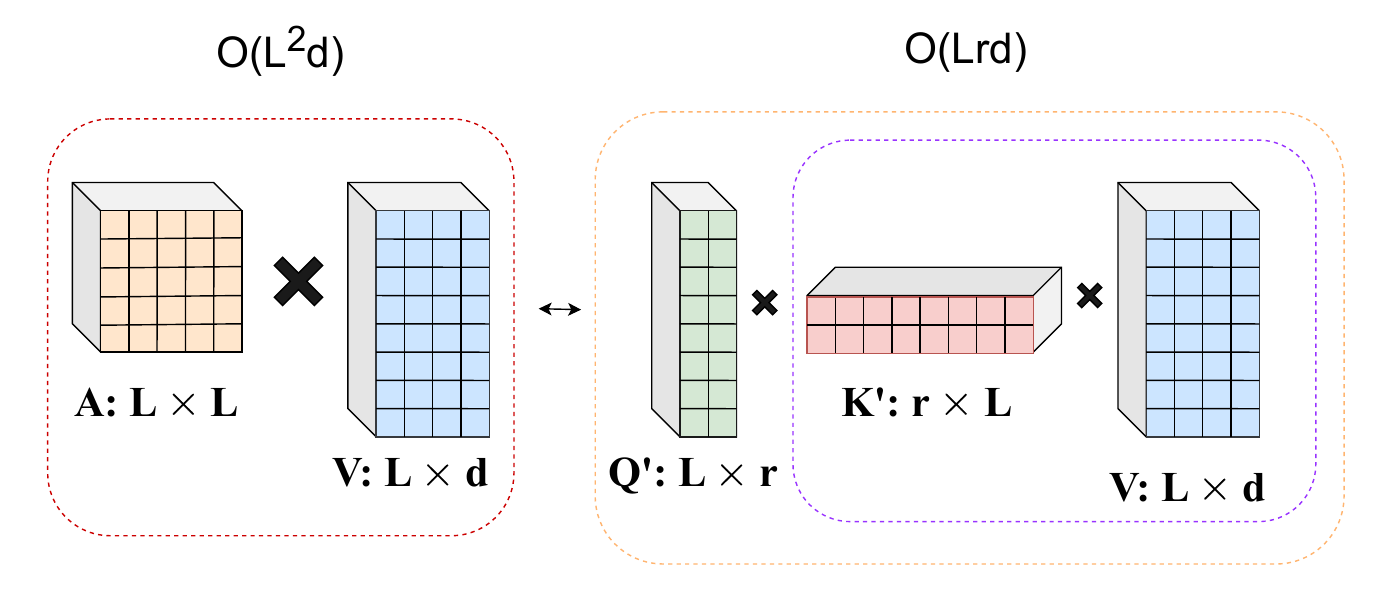}
\caption{Linear attention vs Traditional attention.}
\label{pt}
\end{figure}

In standard transformers, the attention mechanism is defined as:
\begin{equation}
\text{Attention}(Q, K, V) = \text{softmax}\!\left(\frac{QK^T}{\sqrt{d}}\right)V,
\end{equation}
where \( Q \), \( K \), and \( V \) represent query, key, and value matrices, and \( d \) is the key dimension. This formulation has quadratic time complexity relative to sequence length due to the full computation of \( QK^T \), making it impractical for processing the dense sampling required in our application.

Performer achieves linear time complexity by approximating the softmax kernel with random feature maps. Using a random feature map \( \phi(x) \) that satisfies
\begin{equation}
\exp\!\left(x^T y\right) \approx \phi(x)^T \phi(y),
\end{equation}
the attention mechanism can be reformulated by,
\begin{equation}
\mathrm{Att}(Q,K,V)
\approx
D^{-1}\,\phi(Q)\Big(\phi(K)^\top V\Big),
\end{equation}
where $D=\mathrm{diag}\!\Big(\phi(Q)\big(\phi(K)^\top \mathbf{1}_L\big)\Big)$. This factorization eliminates the need to compute the full \( QK^T \) matrix, reducing time complexity from quadratic to linear with respect to sequence length. As Fig.~\ref{pt} shows, $Q' \in \mathbb{R}^{L \times r}$ and $K' \in \mathbb{R}^{r \times L}$. By applying the associativity of matrix multiplication, the computation is first performed on $K'^\top V \in \mathbb{R}^{r \times d}$, followed by multiplication with $Q'$, reducing the total time complexity to $\mathcal{O}(Lrd)$. This is significantly more efficient in cases where $L \gg r, d$, which is strictly followed in our model.

\begin{figure}[htbp]
\centering
\includegraphics[width=\linewidth]{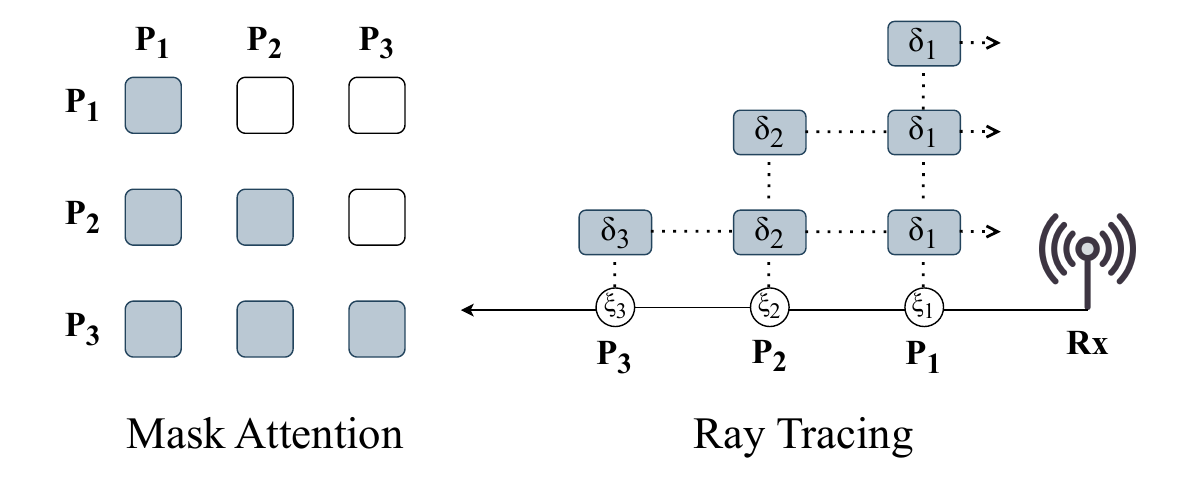}
\caption{Similarity of mask attention and ray tracing.}
\label{com}
\end{figure}

Furthermore, our model employs masked attention to integrate with the ray tracing module. As illustrated in Fig.~\ref{com}, both masked attention and ray tracing propagate information sequentially along a defined path, where each element’s contribution is weighted by its relationship to earlier interactions. In masked attention, a causal mask ensures that each token only attends to preceding tokens, accumulating context step by step. This mechanism parallels ray tracing, where signals are successively attenuated and phase-shifted as they traverse voxels, with downstream values conditioned on upstream interactions. Theoretical justification for this design stems from the causal nature of electromagnetic wave propagation: the field at a point is determined by cumulative contributions from prior interactions governed by Maxwell’s equations. By enforcing directionality and temporal order, masked attention aligns with the principle of causality and preserves phase coherence across interactions. This structured aggregation enables the model to accurately capture long-range dependencies and interference patterns, making masked attention a natural and effective analogy to the cumulative, directional integration process central to ray tracing.

We stack two masked Performer modules to enhance feature integration. The first module captures local dependencies, while the second models broader interactions. This structure enables the network to represent both short-range and long-range relationships along the ray path.

\subsubsection{Feature Integration and Output}
Finally, we employ a linear layer to obtain the high-dimensional attenuation representation $\delta$. This representation is closely related to the physical properties of each voxel, such as material type and scene density, and is independent of the incoming signals. It encodes how electromagnetic waves are attenuated as they pass through the environment, capturing effects including absorption, reflection, and scattering. We feed both the Linear Transformer’s output and the original encoded ray features into this linear layer to ensure stable training and to preserve the original ray feature information.

In addition to the attenuation representation $\delta$, the attenuation network also produces a feature representation \textit{f} that encodes rich information about the ray's interaction with the environment. This feature representation serves as input to the radiance network, enabling end-to-end learning of the complete electromagnetic behavior.

\subsection{The Radiance Network}
The radiance network models electromagnetic signal radiation in environments, sharing a similar structure to the attenuation network. It consists of three components: a PowerMLP module, a Performer module, and a residual connection.

\subsubsection{KAN Module}
The KAN module in the radiance network processes three inputs: the feature representation \textit{f} from the attenuation network, encoded ray direction E($\omega$), where $\omega$ is the direction of the ray, and encoded transmitter position E($P_\text{TX}$) where $P_\text{TX}$ is the location of the transmitter. The feature vector contains information about environmental properties affecting signal propagation. The ray direction encoding provides data on signal propagation angles, important for modeling direction-dependent effects like specular reflection. The transmitter position encoding helps model how signals from different locations interact with the environment. The radiance network only needs to map highly encoded features to a per-voxel emission response. This mapping is relatively simple and well-conditioned by the upstream features, therefore two layers of PowerMLP with wider hidden dimensions is adopted to handle the additional input features.

\subsubsection{Performer and Feature Integration}
The KAN output then passes through a two-layer masked Performer. We include a Performer block in the radiance network because it excels at modeling dependencies between distant points along each ray—dependencies that arise from multipath reflections, diffractions, and other scattering phenomena, while it also refines voxel feature integration.

By using the same number of Transformer layers and architecture as in the attenuation network, we ensure both branches have comparable capacity to learn spatial correlations at similar depths. This architectural alignment simplifies hyperparameter selection, balances gradient propagation across the two networks, and leads to more stable training and consistent feature integration. 

\subsubsection{Residual Connection and Output}
The Performer output enters a linear layer together with the original ray information, encoded ray direction E($\omega$), and encoded transmitter position E($P_\text{TX}$) through a residual connection. This connection preserves essential information and enhances training stability by providing a direct path for gradient flow during backpropagation.

The radiance network produces a direction-dependent RF signal emission representation $\xi$ that is retransmitted from the voxel along a specified direction. This representation encodes how electromagnetic signals are modified when interacting with materials and structures, capturing effects such as reflection, diffraction, and scattering.

\subsection{Ray Tracing}
We model radio wave propagation through a voxel grid using a volumetric approach that mirrors physical scattering. Each voxel is characterized by the attenuation representation $\delta(\mathbf{x})$, which quantifies how much that voxel attenuates the passing signal, and the signal emission representation $\xi(\mathbf{x})$, which is the signal power that the voxel re-radiates in the direction of the ray due to scattering of the incoming RF signal into that direction. As a ray travels through the medium, its intensity $I(s)$ is continuously diminished by attenuation and augmented by scattered energy, where $s$ is the position of a voxel along the ray. This can be described by a differential propagation equation similar to the radiative transfer law:
\begin{equation}
\frac{dI(s)}{ds} = -\,\delta(s)\,I(s) + \xi_s(s),
\end{equation}
where the first term $-\delta(s)I(s)$ represents exponential attenuation of the beam as it travels, while the second term $\xi_s(s)$ adds the radiance contributed by that voxel into the ray (i.e. the scattered signal). Solving the above equation along a ray yields an integral formulation that accumulates all voxel contributions. For a ray from the RX at $s=0$ out to a distance $D$, the received signal intensity $I(0)$ is formulated by:
\begin{equation}
I(0) = \int_{0}^{D} \exp\left(-\int_{0}^{s} \delta(t)dt\right)\,\xi_s(s)\,ds,
\end{equation}
where $\exp\left(-\int_0^s \delta(t),dt\right)$ represents the transmittance from point $s$ to the RX, indicating the fraction of the signal that survives attenuation along the path between $s$ and $0$. In this formulation, each tiny segment $ds$ of the ray contributes an amount of radiance $\xi_s(s)ds$, but that contribution is reduced by the cumulative attenuation of all voxels between that segment and the RX.

In practical applications, measurements typically rely on discrete samples rather than continuous functions. For computational efficiency, we assume uniform distributions along each optical path. Under this assumption, we first apply mean pooling across ray features, allowing us to approximate the integral received signal at the RX as a discrete sum over N voxels (i.e. $i=1,\ldots,N$) along the ray, shown as:
\begin{equation}
I_{\text{RX}} \approx \sum_{i=1}^{N} exp\left( \sum_{j=1}^{i-1} \delta_j \right) \xi_i,
\end{equation}
where $\xi_i$ is the radiance emitted by voxel $i$ toward the RX, and $\delta_j$ is the voxel transmittance, with $\Delta s$ as the ray step. The product $\sum_{j=1}^{i-1}\delta_j$ represents the cumulative attenuation from all voxels between voxel $i$ and the RX. Thus, the signal emission $\xi_i$ of each voxel is scaled by the accumulated attenuation along its path, ensuring that signals from voxels farther from the RX (which in another way, are nearer to the TX) are appropriately reduced by the attenuation caused by intervening voxels.

%% file: sec/4_experiments.tex
\section{Experiments}
\label{sec:exp}
In this section, we describe our datasets and present the results. We also provide explanatory visualizations that demonstrate how our findings align with real-world scenarios.
\subsection{Datasets} 
We employ two types of datasets: realistic and synthetic. For realistic data, we use (1) a MIMO-CSI dataset from the Argos channel dataset~\cite{shepard2016understanding}, which provides extensive wireless measurements with ground truth positions for localization and channel estimation, and (2) an RFID dataset from the NeRF$^2$~\cite{nerf2} repository containing 10,000 spatial spectrums, represented as $360 \times 90$ grids captured from a $4 \times 4$ antenna array at 915~MHz, with 8,000 samples for training and 2,000 for testing. For synthetic data, we adopt the NewRF~\cite{lu2024newrf} repository, which simulates conference and bedroom environments in Matlab. Each scene includes one transmitter and $443$，$975$ randomly placed receivers, respectively, with measurements collected across 52 OFDM subcarriers. We use 80\% of the data for training and 20\% for testing. The model is trained on the first 26 uplink subcarriers and evaluated by predicting the remaining 26 downlink subcarriers.

\subsection{Results}
All results are generated by Optuna~\cite{akiba2019optuna}. Optuna is an automatic hyperparameter optimization framework designed to efficiently search and tune model parameters using state-of-the-art algorithms.

\begin{figure*}[htbp]
  \centering
  \newlength{\threecolwidth}
  \setlength{\threecolwidth}{\dimexpr(\textwidth-1.0\columnsep)/3\relax}
  
  \begin{subfigure}[t]{\threecolwidth}
    \centering
    \includegraphics[width=1\linewidth, height=6.5cm, keepaspectratio]{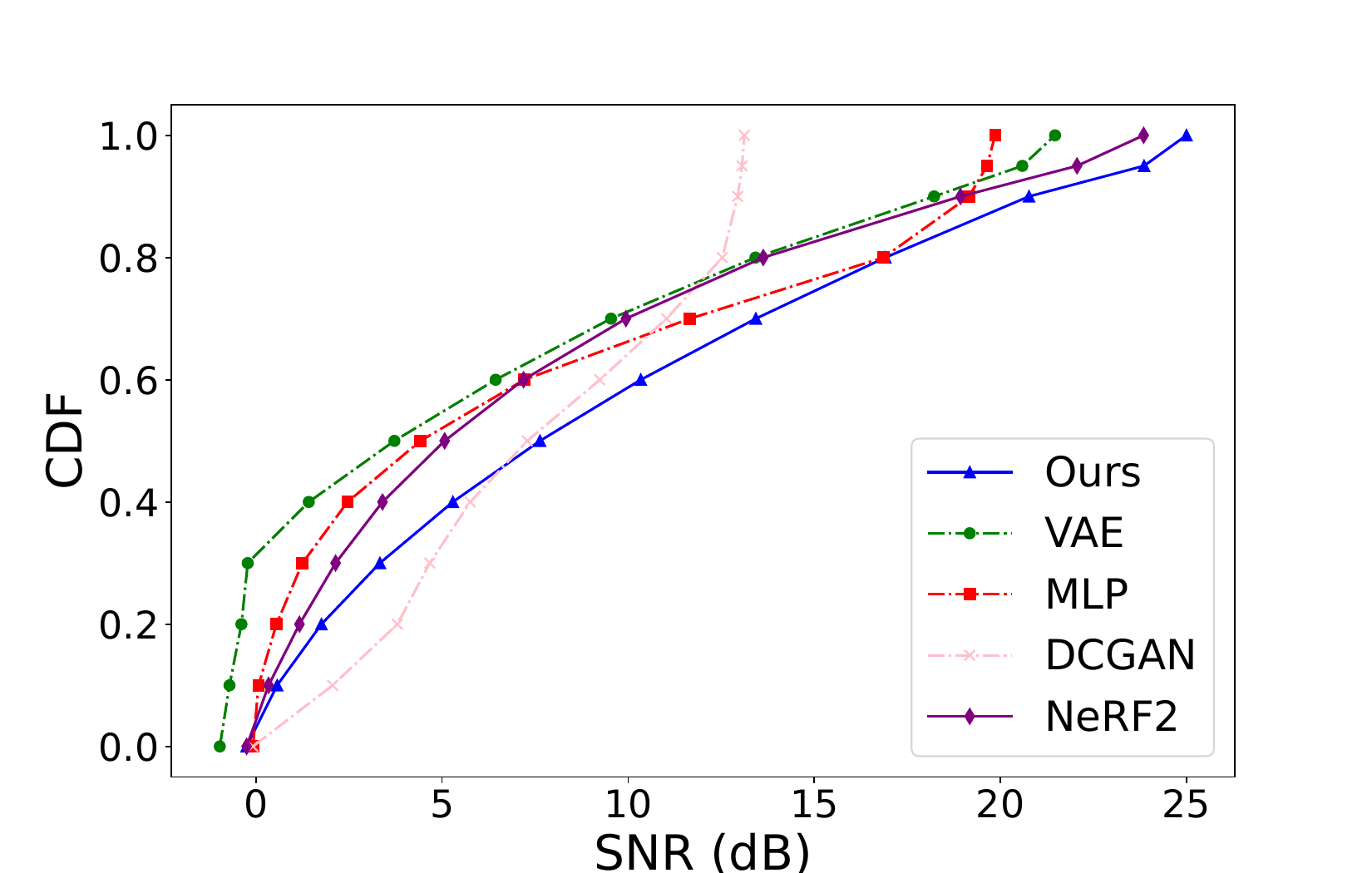}
    \caption{MIMO-CSI dataset.}
    \label{fig:sub1}
  \end{subfigure}\hfill
  \begin{subfigure}[t]{\threecolwidth}
    \centering
    \includegraphics[width=1\linewidth, height=6.5cm, keepaspectratio]{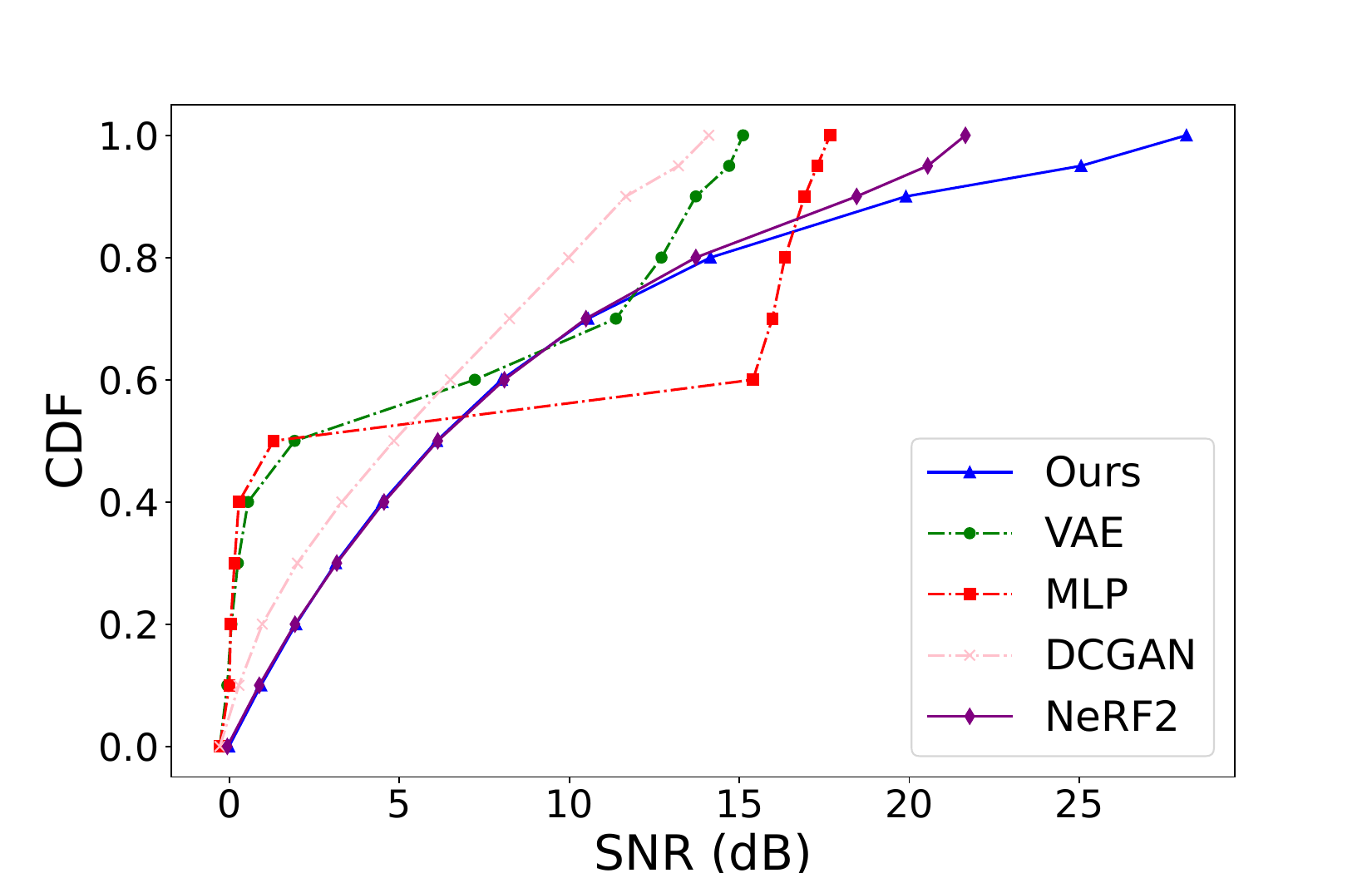}
    \caption{Bedroom dataset.}
    \label{fig:sub2}
  \end{subfigure}\hfill
  \begin{subfigure}[t]{\threecolwidth}
    \centering
    \includegraphics[width=1\linewidth, height=6.5cm, keepaspectratio]{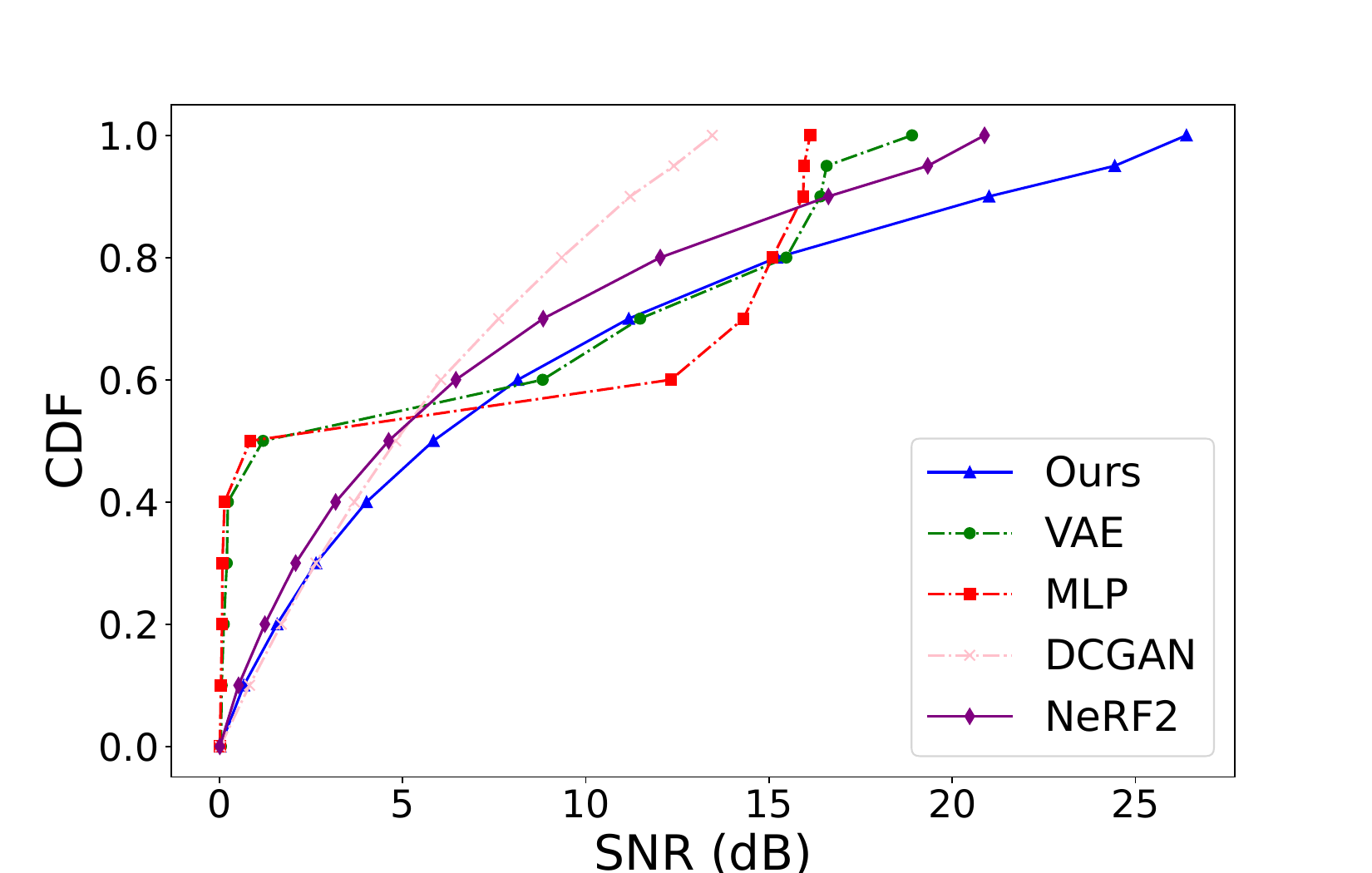}
    \caption{Conference dataset.}
    \label{fig:sub3}
  \end{subfigure}
  \vspace{0.8em}
  \caption{CDF-SNR performance comparison across three datasets.}
  \label{fig:three-wide}
\end{figure*}

\subsubsection{Baselines}
We evaluate our method against four different baseline models. First, we implement a four-layer MLP with hidden units of 256, 128, 64, and 64, followed by a linear output layer. Second, we adapt the VAE architecture from FIRE~\cite{10.1145/3447993.3483275}, using three encoder layers and four decoder layers with identical hidden unit configurations. Third, we develop an MLP-based DCGAN~\cite{radford2015unsupervised} comprising four generator and discriminator layers, starting with a maximum hidden dimension of 512 that decreases by half in each subsequent layer. Finally, we construct NeRF$^2$ following their established architectures. We use signal-to-noise ratio (SNR) and structural similarity index measure (SSIM) as our evaluation metrics. 

\subsubsection{Experiment Results}
Our proposed KANNA-NeRF method demonstrates superior performance across all evaluation metrics compared to baseline approaches. In synthetic data experiments. Table~\ref{tab:performance_s} reports results for five methods on four datasets. The left two columns show the experiment results for realistic datasets. For the MIMO scenario in the synthetic dataset, our evaluation reveals a clear performance hierarchy among the tested methods. KANNA-NeRF achieves the highest SNR at 25.00 . The original NeRF$^2$ demonstrates strong performance with 23.85 dB, while VAE and MLP deliver moderate results at 21.47 dB and 19.86 dB, respectively. DCGAN shows substantially lower performance with only 13.12 dB. In the RFID scenario, a similar pattern emerges with KANNA-NeRF leading at 0.86 SSIM, while the SSIM for NeRF$^2$, VAE, MLP, and DCGAN are 0.82, 0.73, 0.71, and 0.52, respectively.

\begin{table}[t]
    \centering
    \renewcommand{\arraystretch}{1.2}
    \setlength{\tabcolsep}{4pt}
    \small
    \begin{tabularx}{\linewidth}{X|cc|cc}
        \toprule
        \multirow{2}{*}{\textbf{Method}} & \multicolumn{2}{c|}{\textbf{Realistic datasets}} & \multicolumn{2}{c}{\textbf{Synthetic datasets}} \\
        \cmidrule{2-5}
        & \begin{tabular}[c]{@{}c@{}}\textbf{MIMO}\\(\textbf{SNR/dB$\uparrow$})\end{tabular} & 
          \begin{tabular}[c]{@{}c@{}}\textbf{RFID}\\(\textbf{SSIM $\uparrow$})\end{tabular} & 
          \begin{tabular}[c]{@{}c@{}}\textbf{Bedroom}\\(\textbf{SNR/dB$\uparrow$})\end{tabular} & 
          \begin{tabular}[c]{@{}c@{}}\textbf{Conference}\\(\textbf{SNR/dB$\uparrow$})\end{tabular} \\
        \midrule
        MLP & 19.86 & 0.71 & 17.67 & 16.12 \\
        VAE & 21.47 & 0.73 & 15.11 & 18.90 \\
        DCGAN & 13.12 & 0.52 & 14.10 & 13.45 \\
        NeRF$^2$ & 23.85 & 0.82 & 21.65 & 20.88 \\
        \midrule
        \textbf{KANNA-NeRF (Ours)} & \textbf{25.00} & \textbf{0.86} & \textbf{28.15} & \textbf{26.39} \\
        \bottomrule
    \end{tabularx}
    \caption{Comparison of KANNA-NeRF against baseline methods on realistic and synthetic datasets.}
    \label{tab:performance_s}
\end{table}

The remaining two columns show the results of synthetic scenes. In the Bedroom environment, KANNA-NeRF again demonstrates superior performance with an SNR of 28.15 dB, establishing a clear advantage over the baseline NeRF$^2$ method at 21.65 dB, while traditional approaches show considerably lower performance: MLP reaches 17.67 dB, VAE declines to 15.11 dB, and DCGAN performs poorest at 14.10 dB. The Conference environment results reinforce this pattern, with KANNA-NeRF achieving the highest SNR of 26.39 dB. The performance gap widens further for the remaining methods, with NeRF$^2$ at 20.88 dB, VAE at 18.90 dB, MLP at 16.12 dB, and DCGAN again showing the lowest performance at 13.45 dB. The superior performance of our KANNA-NeRF method compared to other approaches can be attributed to several architectural advantages. Additionally, Fig.~\ref{fig:fig4} shows the prediction example of our model in the bedroom dataset. The yellow uplink CSI data, containing phase and amplitude measurements across 26 subcarriers, serves as input for our method. The blue downlink CSI predictions align closely with the red ground truth values. This near-perfect overlap validates our model's predictive accuracy.

Based on Table~\ref{tab:performance_s}, traditional methods such as MLP, VAE, and DCGAN show a large performance gap compared with NeRF-based models, likely due to their limited spatial understanding. While NeRF$^2$ leverages volumetric representation to capture spatial relationships, it struggles to model voxel interactions and fine-grained feature dependencies. Our KANNA-NeRF addresses these limitations through a ray-in-ray-out structure enhanced with attention and KAN mechanisms, providing stronger context awareness and feature correlation across spatial dimensions, and thus achieving more accurate reconstructions in complex real-world scenarios.

\begin{figure}[htbp]
    \centering
    \setlength{\tabcolsep}{1pt} 
    \renewcommand{\arraystretch}{0.5} 
    \begin{tabular}{ccc}
        \includegraphics[width=0.30\linewidth]{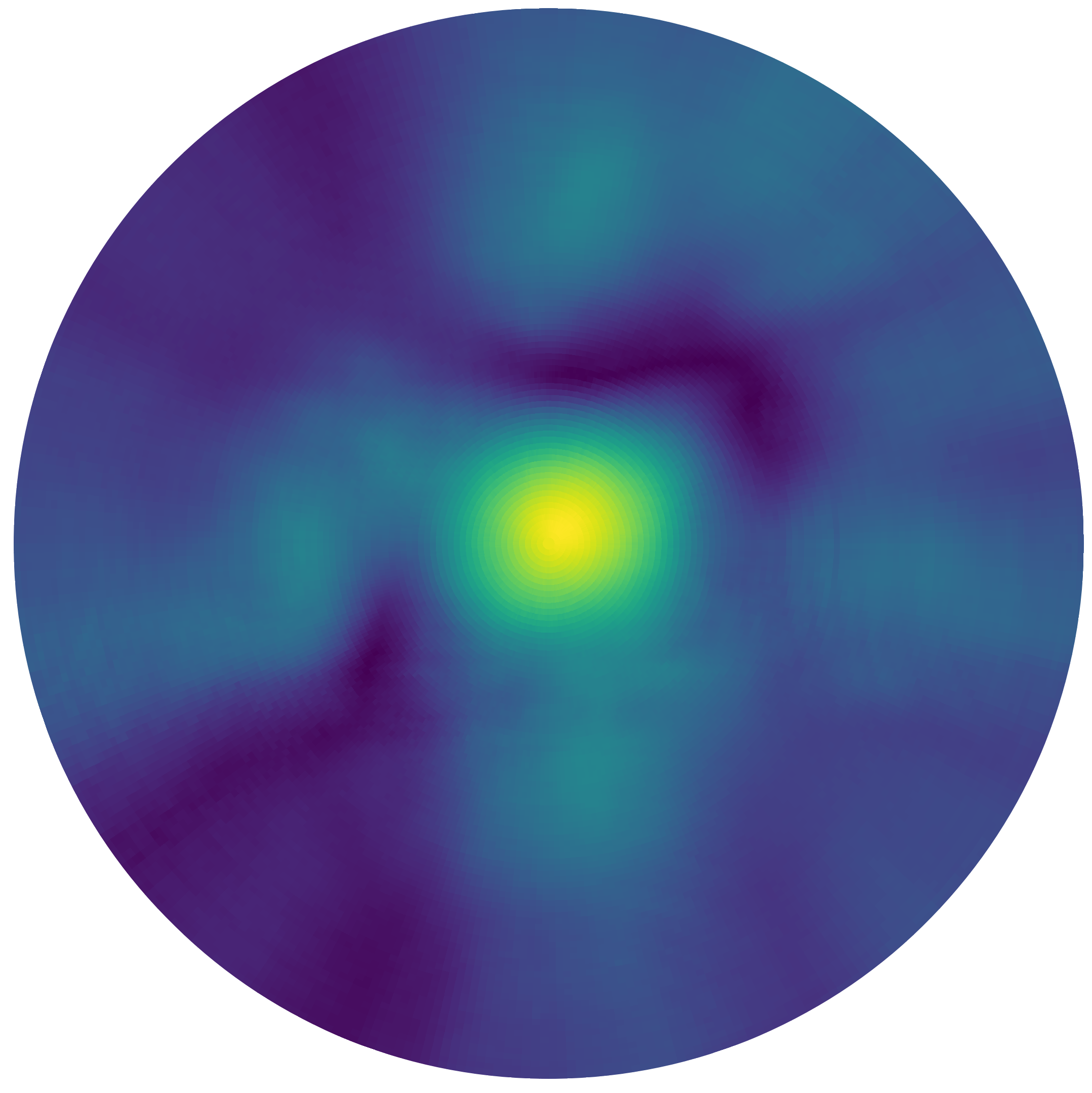} & 
        \includegraphics[width=0.30\linewidth]{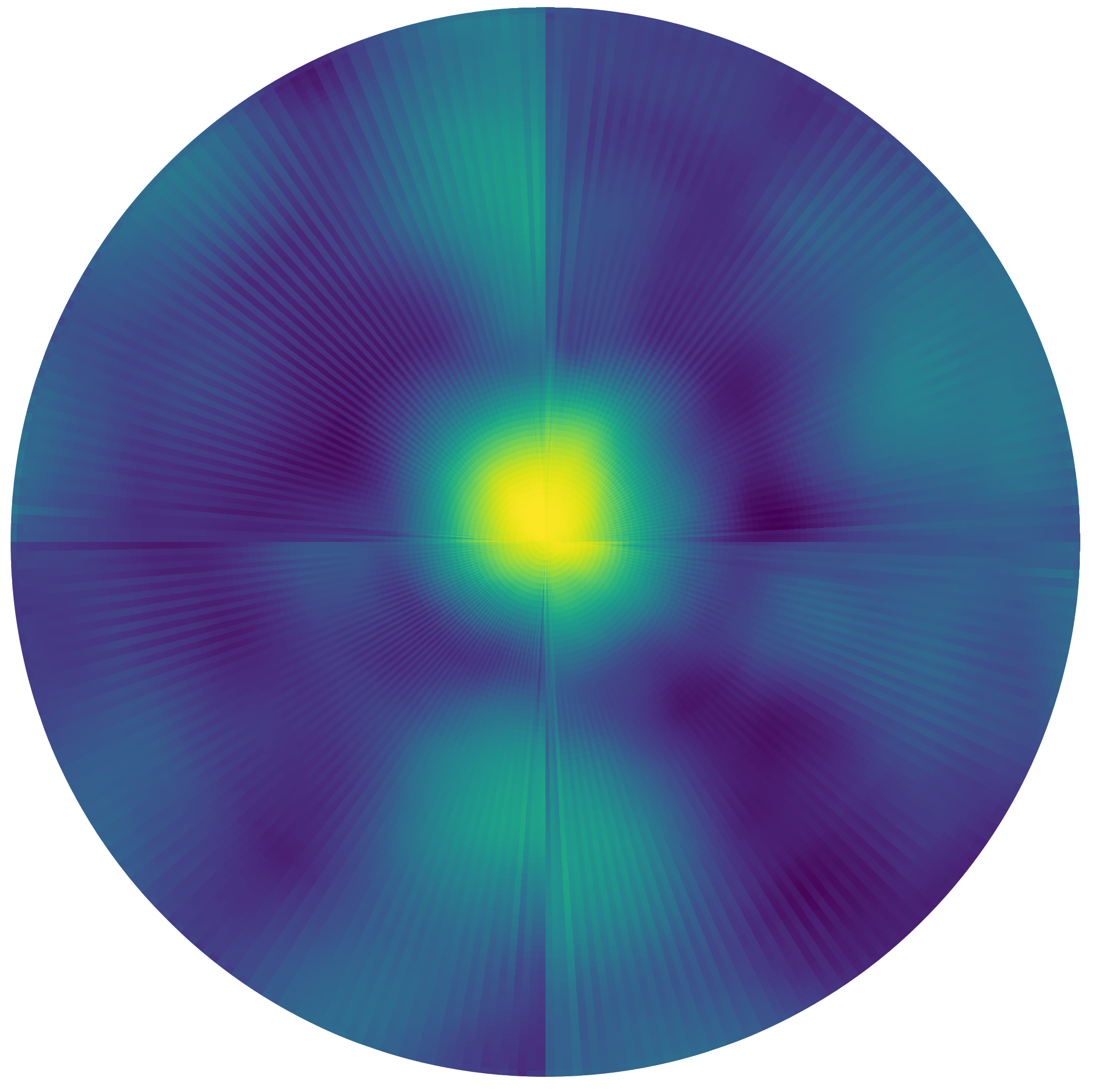} &
        \includegraphics[width=0.30\linewidth]{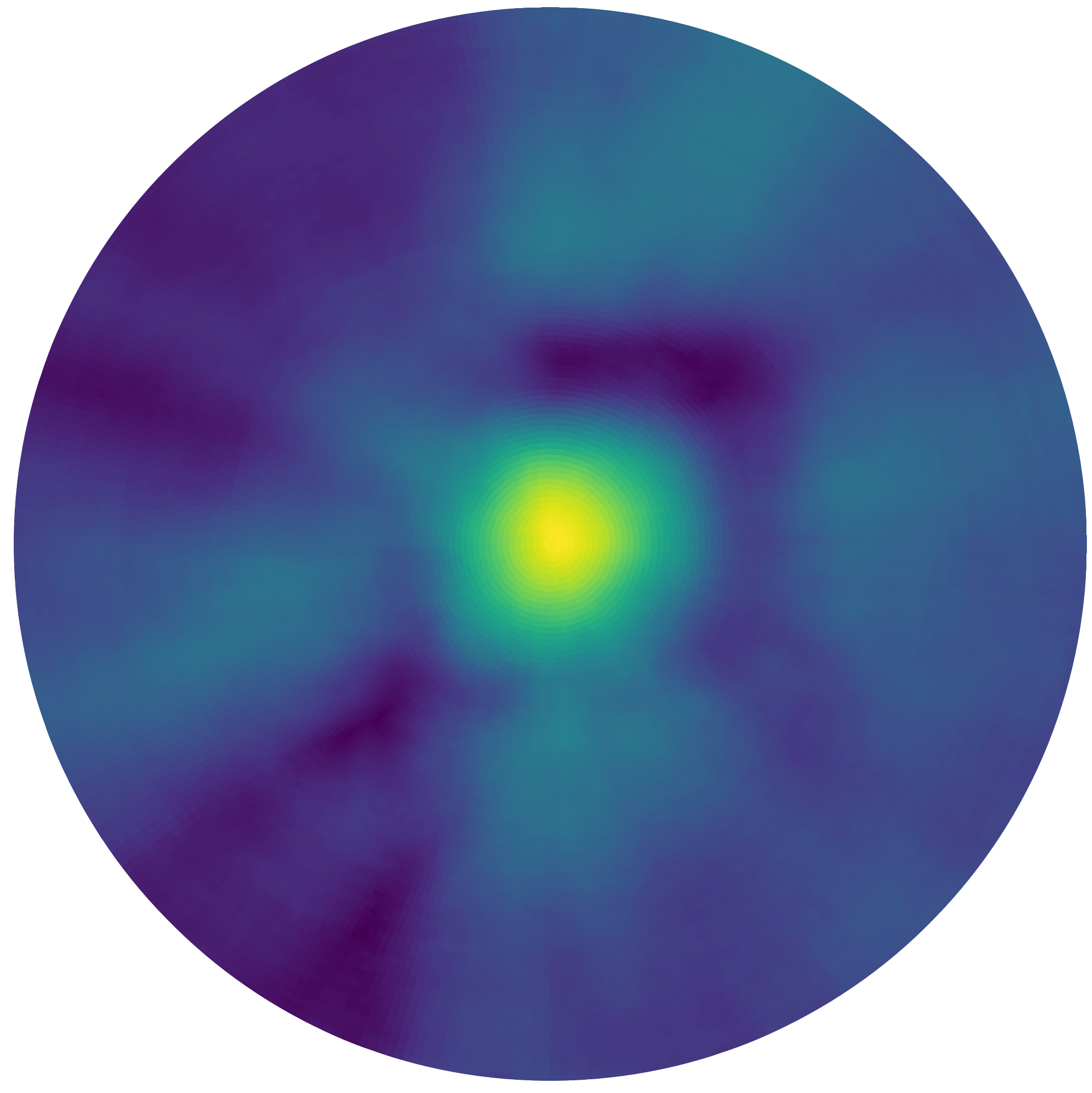} \\
        (a) Ours & (b) DCGAN & (c) NeRF$^2$ \\[2mm]
        \includegraphics[width=0.30\linewidth]{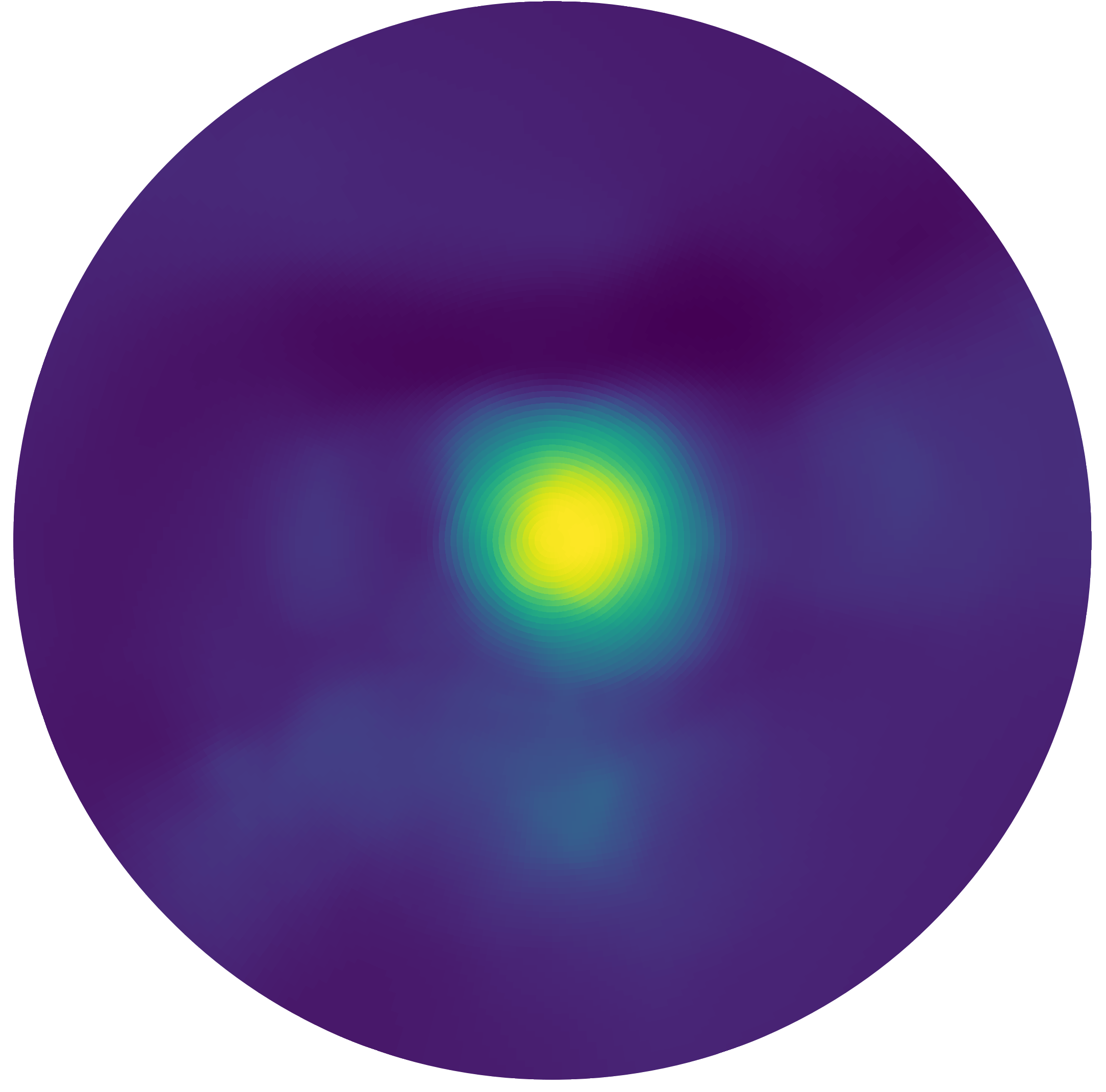}&
        \includegraphics[width=0.30\linewidth]{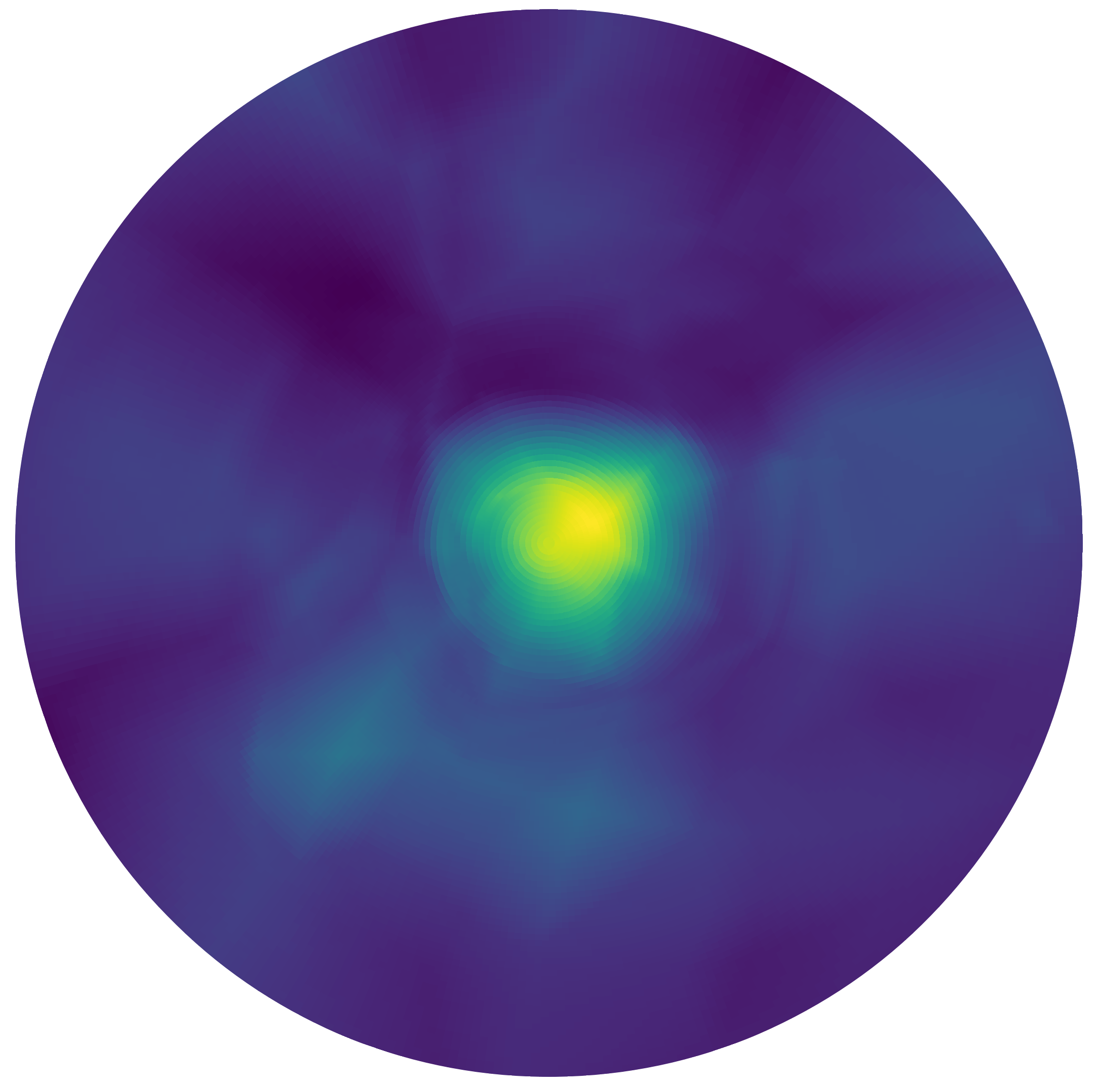}&
        \includegraphics[width=0.30\linewidth]{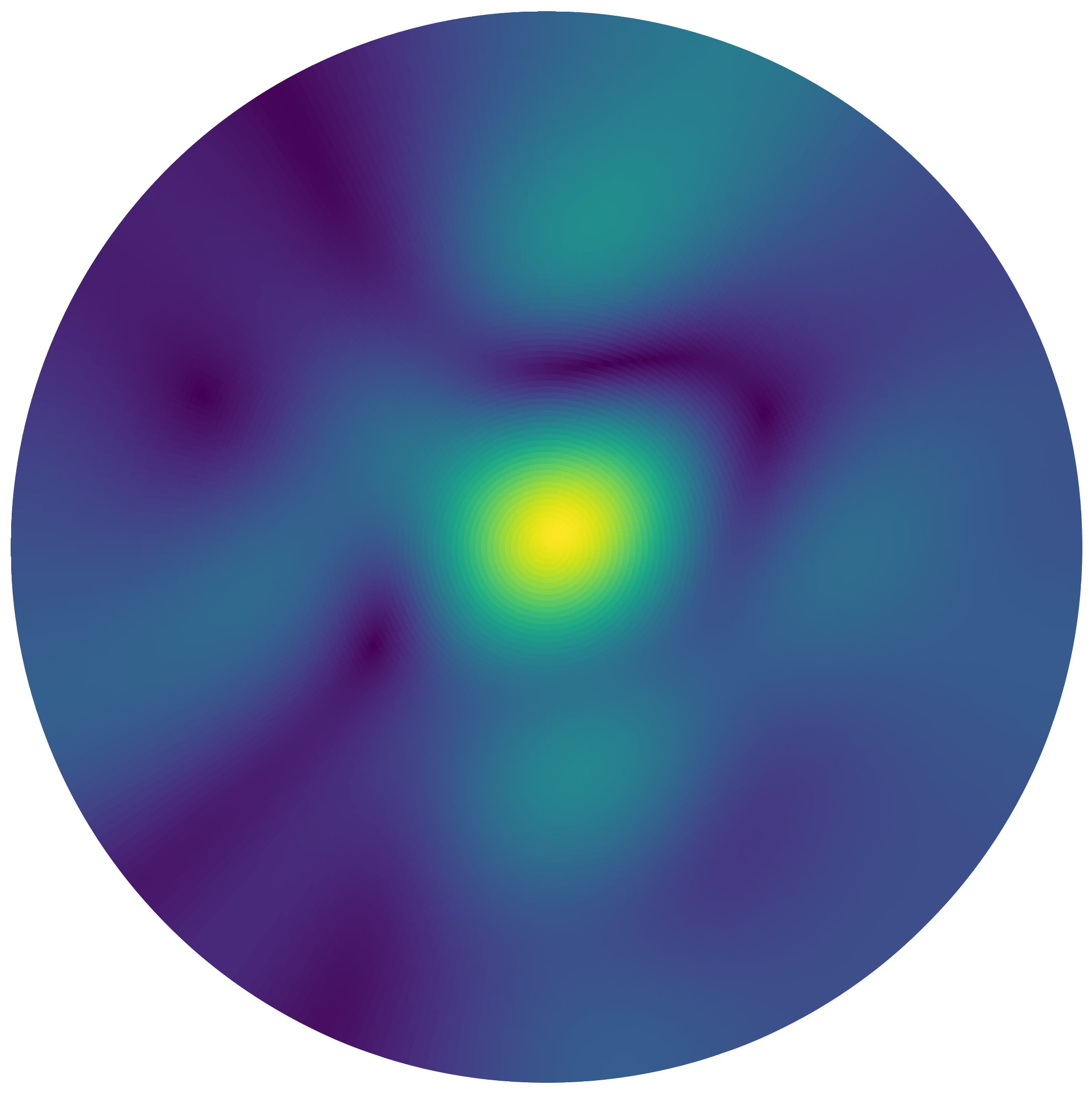} \\
        (d) MLP & (e) VAE & (f) Label \\
    \end{tabular}
    \caption{2D spatial spectrum visualizations.}
    \label{fig:your_label}
\end{figure}

Additionally, we present the Cumulative Distribution Function (CDF) of the SNR between predicted received signals and ground truth for three CSI-based datasets in Fig.~\ref{fig:three-wide}. For each test case across all methods and datasets, we calculate individual SNR values to measure reconstruction quality. These values are sorted in ascending order to create a cumulative distribution, where each point represents the proportion of test cases achieving that SNR or lower. This CDF-SNR curve effectively displays the complete distribution of results, showing performance consistency across various conditions. Better performance is indicated by curves that rise more steeply and extend further rightward. Our results demonstrate superior performance as our curve consistently dominates across the entire SNR range.

\begin{figure}[htbp]
    \centering
    \includegraphics[width=0.95\linewidth]{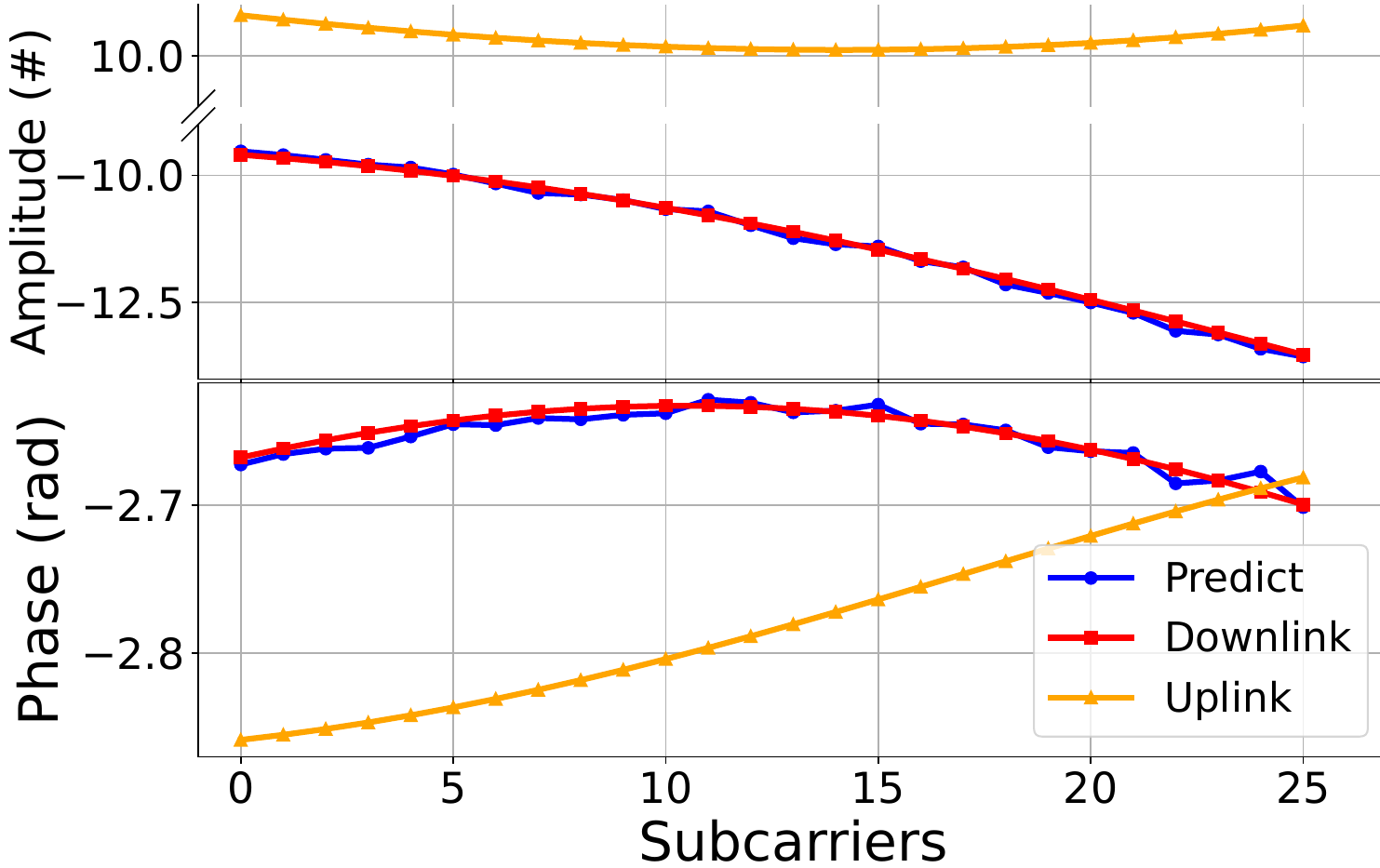}
    \caption{Channel Amplitude \& Phase in the bedroom dataset.}
    \label{fig:fig4}
\end{figure}

\renewcommand{\arraystretch}{1.33}
\captionsetup[table]{skip=10pt} 
\begin{table}[ht]
    \centering
    \footnotesize
    \begin{tabular}{|p{2.7 cm}|c|c|c|c|}
    \hline
    & \textbf{Bedroom} & \textbf{Conference} & \textbf{MIMO-CSI} \\
    & \textbf{(SNR / dB$\uparrow$)} & \textbf{(SNR / dB$\uparrow$)} & \textbf{(SNR / dB$\uparrow$)}\\
    \hline
    Ours & \textbf{28.15} & \textbf{26.39} & \textbf{25.00}\\
    Ours (w/o Residual Connection) & 27.30 & 25.44 & 22.74\\
    Ours (w/o Performer) & 24.64 & 24.20 & 22.12\\
    Ours (w/o Data Preprocessing) & 22.31 & 23.00 & 22.53\\
    Ours (w BatchNorm) & 22.22  & 25.49 & 22.03\\
    Ours (w LayerNorm) & 23.88 & 25.08 & 21.56\\
    \hline
    \end{tabular}
    \caption{Ablation study on KANNA-NeRF.}
    \label{tab:abla}
\end{table}

Furthermore, Fig.~\ref{fig:your_label} presents the visualizations of spatial spectrums in the RFID-spectrum dataset over various methods at three positions. These spectrums show how the RX collects signals from various directions. Our method yields predictions that most closely match the ground truth.

\subsubsection{Ablation study}

Table~\ref{tab:abla} presents the ablation study results for our KANNA-NeRF model across three different datasets. In the bedroom dataset, our model achieves the highest SNR of 28.15 dB. If we change the key components in our method, it would result in performance degradation: without residual connections yielding 27.30 dB, without Performer attention mechanism producing 24.64 dB, and without data pre-processing dropping to 22.31 dB. Similarly, adding normalization strategies with BatchNorm or LayerNorm reduced performance to 22.22 dB and 23.88 dB, respectively. The conference dataset exhibits comparable trends with our full model achieving SNR of 26.39 dB, outperforming variants without residual connections at 25.44 dB, without Performer at 24.20 dB, without data pre-processing at 23.00 dB, with BatchNorm at 25.49 dB and with LayerNorm at 25.08 dB, respectively. The MIMO-CSI dataset results further validate our architectural choices, with our method attaining highest SNR of 25.00 dB compared to lower performance when key components are modified: 22.74 dB without residual connections, 22.12 dB without Performer, 22.53 dB without data pre-processing, 22.03 dB with BatchNorm, and 21.56 dB with LayerNorm. The performance degradation observed in wireless datasets when applying normalization techniques may be primarily attributed to the disruption of the KAN layers’ adaptive activation functions. BatchNorm and LayerNorm alter the feature scales and shifts that KAN layers are designed to learn, which may weaken the network’s ability to model complex data relationships and lead to reduced performance.

\begin{figure}[htbp]
    \centering
    \includegraphics[width=\linewidth]{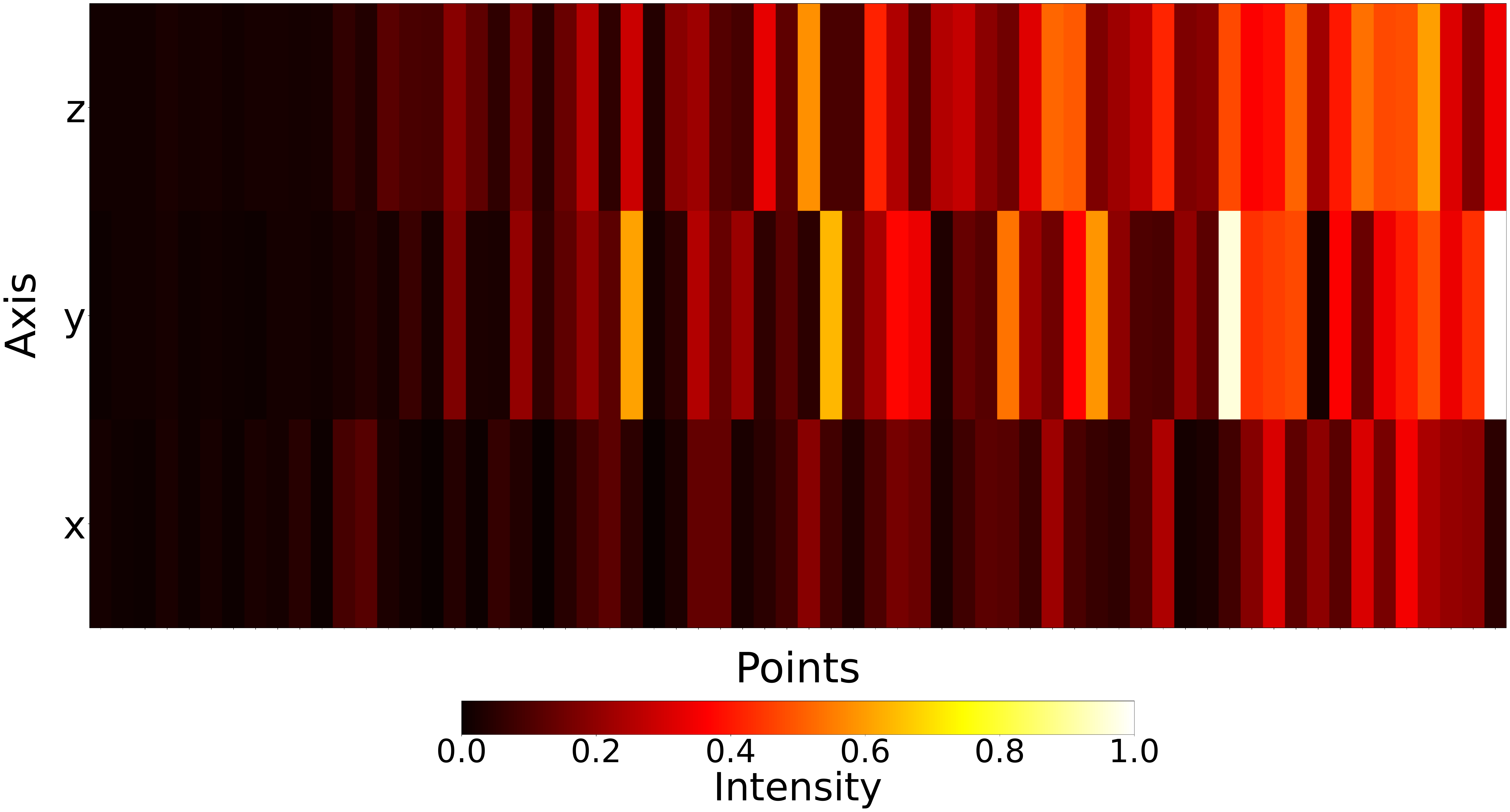}
    \caption{Explanation on sampled points in a ray.}
    \label{fig:overall}
\end{figure}
In this section, we explain our results, focusing on the KAN module's effectiveness. Fig.~\ref{fig:overall} displays the Jacobian matrix with sampled 3D voxel positions, where the y-axis shows 3D positions (x, y, z) and the x-axis represents points sampled along the ray. In wireless spatial signal propagation, the Jacobian matrix indicates how spatial coordinate changes affect the signal field. High values in the matrix highlight regions of rapid signal variation, potentially indicating re-transmit sources. This illustration provides geometric features of the scene, offering essential information about the surrounding environment.
\begin{figure}[htbp]
    \centering
    \includegraphics[width=0.86\linewidth]{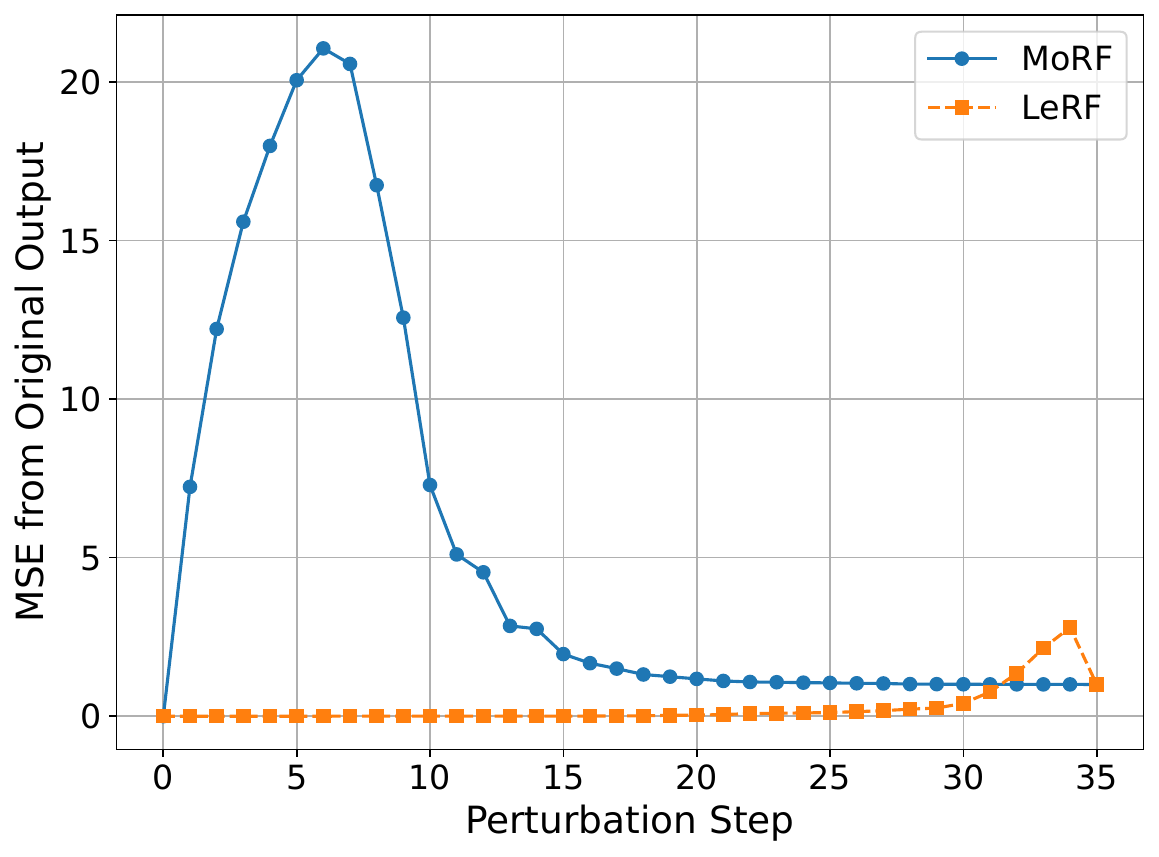}
    \caption{MoRF vs LeRF curves.}
    \label{fig:exp}
\end{figure}

We then present two complementary approaches to validate our assumption: Most Relevant First (MoRF) and Least Relevant First (LeRF). These methods help identify which features contribute most to model decisions shown in Fig.\ref{fig:exp}. Samek et al.\cite{samek2017evaluating} formalized these methods within a framework for assessing neural network explanations. MoRF removes important features first to show how performance degrades, while LeRF removes less important features first to demonstrate model resilience. We determine feature importance by taking absolute values from the Jacobian matrix and summing across output dimensions. We then progressively mask features in either MoRF or LeRF order and measure the MSE between normalized outputs to quantify distortion. Our sampling combines log-spaced and percentile linear methods to capture both early feature importance and comprehensive trends.

Fig.~\ref{fig:exp} shows our analysis results. The MoRF curve rises sharply at first, peaks when critical features are removed, then declines as only minor features remain. This pattern confirms the model's dependence on core features. The LeRF curve stays nearly flat for the first 80\% of steps, showing minimal impact when removing less important features. It demonstrates the model's robustness to non-essential input changes. The LeRF curve only increases once most irrelevant features are masked. The modest final rise occurs because the model output has already approached its baseline. The features identified as important match those highlighted in Fig.~\ref{fig:overall}.

%% file: sec/5_conclusion.tex
\section{Conclusion}
\label{sec:conclu}

In this paper, we present a novel ray-in-ray-out encoder that enhances wireless channel modeling beyond traditional point-to-point approaches. By incorporating contextual information between ray points and implementing masked attention mechanisms, our model effectively captures the complex relationships in wireless propagation environments. The integration of KAN modules instead of conventional MLPs significantly improves the representational capacity of our system, allowing for more accurate signal predictions. Our experimental results demonstrate superior performance across both realistic and synthetic wireless scenarios, consistently outperforming baseline methods. Additionally, our explanatory visualizations clearly link model outputs to physical propagation phenomena. It demonstrates that our method provides more comprehensive and accurate wireless channel modeling.

\section*{Acknowledgments}
This work is supported in part by the U.S. NSF under grants (CNS-2415209, CNS-2321763, CNS-2317190, IIS-2306791, and CNS-2319343).